\documentclass[twocolumn,showpacs,preprintnumbers,amsmath,amssymb,prb,superscriptaddress,aps,10pt]{revtex4-1}

\usepackage{graphicx}

\begin{document}

\title{FCS of superconducting tunnel junctions in non-equilibrium}
\author{H. Soller}
\affiliation{Institut f\"ur Theoretische Physik,
Ruprecht-Karls-Universit\"at Heidelberg,\\
 Philosophenweg 19, D-69120 Heidelberg, Germany}
\date{\today}

\begin{abstract}
We analyse the full counting statistics (FCS) of a superconducting junction in non-equilibrium in the limit of small interface transparency. In this limit we treat both supercurrent and multiple Andreev reflections on equal footing and show how to generalise previous results for both phenomena. Furthermore, we also allow for different gaps of both superconductors and investigate the intermediate regime which allows to make contact with previous results on normal-superconductor heterostructures. We also compare our predictions in this regime to experimental data.
\end{abstract}


\maketitle

\section{Introduction}
The charge transfer statistics of superconducting tunnel junctions have been an object of interest for many years.\cite{PhysRevLett.87.197006,PhysRevLett.91.187001,PhysRevB.70.214512,PhysRevLett.91.187002} At energies below the gap two phenomena are observed: on the one hand, coherent supercurrent of electrons between the superconductors depending on the phase difference\cite{PhysRevLett.87.197006} and on the other hand multiple Andreev reflection (MARs).\cite{PhysRevLett.91.187001,PhysRevLett.91.187002} Andreev reflection (AR) refers to the retroreflection of one electron from another superconductor as a hole leaving a Cooper pair behind. Such processes may happen multiple times if the two leads attached are both superconductors.\\
The properties of Andreev reflection and supercurrent in multiple hybrid systems have been investigated experimentally both for tunnel contacts\cite{PhysRevLett.73.2611,PhysRevLett.75.3533,PhysRevB.86.155402} and more involved geometries involving quantum dots.\cite{0957-4484-22-26-265204}\\
So far a complete analysis of the charge transfer statistics through superconducting junctions has been done either for the dc-component\cite{PhysRevLett.91.187001} or the ac-component in equilibrium.\cite{PhysRevLett.87.197006} Only the current has been analysed completely\cite{PhysRevB.54.7366} but a complete discussion of noise\cite{PhysRevLett.82.4086,PhysRevLett.94.086806} and possible higher cumulants of the charge transport statistics so far is missing.\\
In this work we want to calculate the full counting statistics (FCS), which allows to find the cumulant generating function (CGF) of charge transfer. This function allows to calculate all cumulants of the current flow. However, in order to be able to rationalize the result for the FCS we will restrict ourselves to the case of small transmission. Such restriction will allow for a reasonable truncation of the MAR processes as will be discussed in more detail below. This case will be presented in full detail, especially also incorporating the possibility of different gaps for the two superconductors (SCs).\\
In Section \ref{sectionfcs} we will present the general model and framework we use to calculate the cumulant generating function. Its actual form will be discussed in two steps: in Section \ref{sectionjosephson} we will discuss the properties of supercurrent and in Section \ref{sectionmar} we will present the results for multiple Andreev reflections. For different gaps of the two superconductors there is an intermediate voltage regime which will be analysed in Section \ref{sectionintermediate}. We will compare our results to experimental data in Section \ref{sectionexp} and conclude in Section \ref{sectionconclude}.

\section{Full counting statistics} \label{sectionfcs}

\begin{figure}[ht]
\includegraphics[width=0.7\columnwidth]{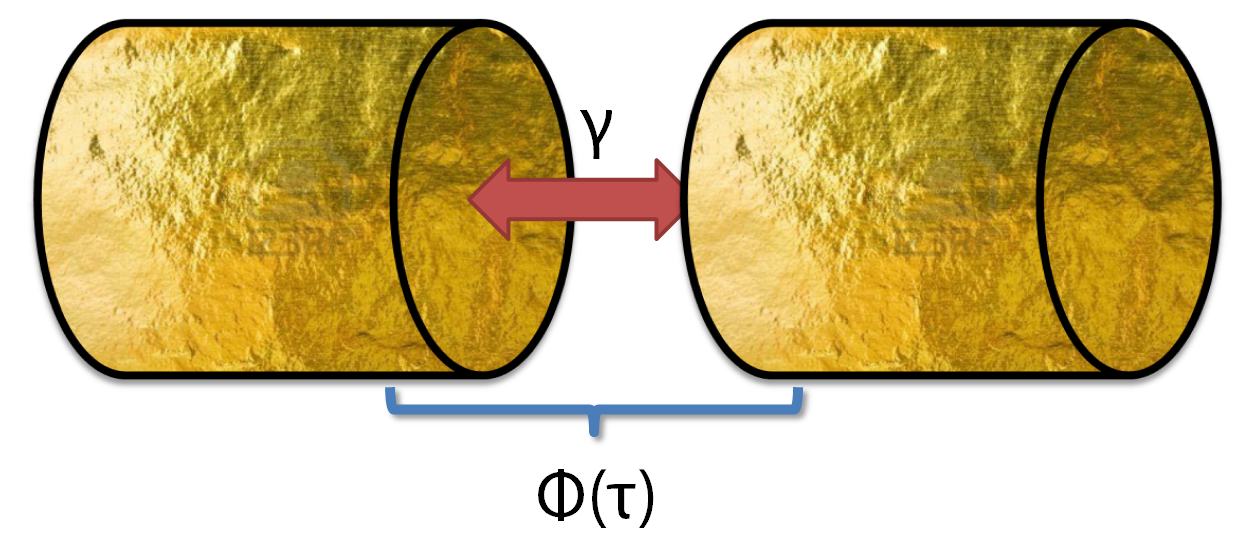}
\caption{Sketch of the system under investigation: two superconductors are tunnel coupled with a tunnel amplitude $\gamma$. Additionally they have a time-dependent phase difference $\phi(\tau) = \phi_0 + 2eV \tau$.}
\label{fig1}
\end{figure}

We calculate the cumulant generating function $\chi(\lambda, \phi)$ using the generalized Keldysh technique\cite{nazarov,PhysRevB.73.195301} which allows to proceed via the Hamiltonian formalism.\cite{PhysRevB.54.7366} $\lambda(t)$ refers to the counting field and $\phi$ is the phase difference of the SCs.\\
$\chi(\lambda,\phi)$ in the case of charge transport is given by
\begin{eqnarray}
\chi(\lambda, \phi) = \sum_q e^{iq\lambda} P_q(\phi) \label{genfou},
\end{eqnarray}
where $P_q(\phi)$ is the probability for the charge $q$ to be transferred through the system during a given (long) measurement time $\tau$.\\
Partial derivatives of $\chi(\lambda, \phi)$ with respect to $\lambda$ give direct access to the cumulants (irreducible moments). Modelling the contact between two SCs as illustrated in Fig. \ref{fig1} is straightforward\cite{PhysRevB.54.7366}
\begin{eqnarray}
H(\tau) &=& H_L + H_R + H_T, \label{hsystem}
\end{eqnarray}
where ($\alpha = L,R$)
\begin{eqnarray*}
H_\alpha = \sum_{k,\sigma} \epsilon_k \alpha_{k\sigma}^+ \alpha_{k\sigma} + \Delta_\alpha \sum_k (\alpha_{k\uparrow}^+ \alpha_{-k\downarrow}^+ + \alpha_{-k\downarrow} \alpha_{k\uparrow}),
\end{eqnarray*}
are the Hamiltonians for the two SCs using units such that $e = \hbar = k_B = 1$. The voltage is applied symmetrically so that $\mu_L = - \mu_R = V/2$ follows for the chemical potentials. The corresponding normal and anomalous Green's functions (GFs) are given in [\onlinecite{PhysRevB.80.184510}] and are abbreviated as $g_{\alpha \sigma}(\omega)$ and $f_\alpha(\omega)$, respectively. The quasiparticle DOS is strongly energy-dependent $\rho_{L/R} = \rho_{0,L/R} |\omega| / \sqrt{\omega^2 - \Delta_{L/R}^2}$.
\\
The tunnel Hamiltonian $H_T$ has to take into account the time-dependent phase of the SCs
\begin{eqnarray}
H_T &=& \gamma \sum_\sigma (e^{i \phi(\tau)/2} L_\sigma^+ R_\sigma + e^{-i \phi(\tau)/2} R_\sigma^+ L_\sigma), \label{htunnel}
\end{eqnarray}
where $\gamma$ is the amplitude of the tunneling coupling and $L_\sigma, \; R_\sigma$ refer to the field components at $x=0$ where tunneling is assumed to occur.\\
To study the FCS we calculate the CGF as a generalized Keldysh partition function. The connection to the system's Hamiltonian in Eq. (\ref{hsystem}) is given by\cite{PhysRevB.70.115305}
\begin{eqnarray}
\ln \chi(\lambda, \phi, \tau) = \left \langle T_{\cal C} \exp \left(-i \int_{\cal C} T^{\lambda(t)} dt\right) \right \rangle,
\end{eqnarray}
where $T^{\lambda(t)}$ denotes Eq. (\ref{htunnel}) after the substitution $R_\sigma \rightarrow R_\sigma e^{-i \lambda(t)/2}$. ${\cal C}$ means the Keldysh contour and $T_{\cal C}$ means time ordering on it. The measuring field $\lambda(t)$ has to be both time- and contour-dependent. It changes sign on the different branches of the Keldysh contour to account for the charge transfer. Additionally $\lambda(t)$ is nonzero only during the time interval $[0, \tau]$. We use the standard expression\cite{PhysRevB.73.195301}
\begin{eqnarray}
\frac{\partial \ln \chi(\lambda, \phi, \tau)}{\partial \lambda} = - i \tau \left \langle T_{\cal C} \frac{\partial T^{\lambda(t)}}{\partial \lambda}\right\rangle_\lambda, \label{lderiv}
\end{eqnarray}
to find the CGF as the counting field derivative of $T^\lambda$. Compared to the case of tunnel contacts between normal metals and SCs\cite{PhysRevB.50.3982,PhysRevLett.87.067006,Soller2011425,PhysRevB.85.174512} the counting field derivative in Eq. (\ref{lderiv}) not only has normal but also anomalous contributions (due to superconductivity) leading to
\begin{eqnarray}
&& \left \langle \frac{\partial T^{\lambda(t)}}{\partial \lambda}\right\rangle_\lambda = - \frac{\gamma^2}{2} \sum_\sigma \int \frac{d\omega}{2\pi} \nonumber\\
&& \left[\underbrace{e^{i \lambda} g_{L\sigma}^{-+}(\omega) {\cal G}_{R\sigma}^{\lambda +-}(\omega) - e^{-i \lambda} g_{L\sigma}^{+-}(\omega) {\cal G}_{R\sigma}^{\lambda-+}(\omega)}_{\mbox{normal contribution}} \right. \nonumber\\
&& \left. \underbrace{+ e^{-i \lambda} (f_L^+)^{+-}(\omega) {\cal F}_R^{\lambda-+}(\omega) - e^{i \lambda} f_L^{-+}(\omega) ({\cal F}_R^{\lambda+})^{+-}(\omega)}_{\mbox{anomalous contribution}} \right],\nonumber\\
&&  \label{derivfull}
\end{eqnarray}
where ${\cal G}$ and ${\cal F}$ refer to the exact-in-tunneling and $\lambda$-dependent GFs and indeces $\pm \pm$ refer to the positions of the two time arguments on the Keldysh contour.\\
Eq. (\ref{derivfull}) can then be integrated with respect to $\lambda$ to access the CGF. The normal contribution gives rise to MARs and quasiparticle tunneling, whereas the anomalous contribution gives rise to Josephson tunneling. We discuss both parts separately.

\section{Josephson tunneling} \label{sectionjosephson}

We first evaluate the second part of the expression in Eq. (\ref{derivfull}), meaning
\begin{eqnarray*}
\left \langle \frac{\partial T_{a,1}^{\lambda(t)}}{\partial \lambda}\right\rangle_\lambda &=& - \frac{\gamma^2}{2} \sum_\sigma \int \frac{d\omega}{2\pi} \left(e^{-i \lambda} (f_L^+)^{+-}(\omega) {\cal F}_R^{\lambda-+}(\omega) \right. \nonumber\\
&& \left. - e^{i \lambda} f_L^{-+}(\omega) ({\cal F}_R^{\lambda+})^{+-}(\omega)\right).
\end{eqnarray*}
As discussed before we can evaluate ${\cal F}_L$ and ${\cal F}_R$ exactly by means of their corresponding Dyson equation.\cite{PhysRevB.54.7366} However, the result is more transparent if we only give the first order in $\Gamma = \pi^2 \gamma^2 \rho_{0L} \rho_{0R}$, keeping in mind that this is reasonable only for $T \ll \Delta_{L/R}$. In this case we can use the approximation of the GFs as in [\onlinecite{PhysRevLett.80.2913}], where the tunneling self-energy is real and purely off-diagonal for energies below the gap and diagonal for energies above the gap. The result for the CGF is then
\begin{eqnarray}
&&\ln \chi_{a,1} (\lambda, \phi, \tau)\nonumber\\
&=& 2 \tau \int \frac{d\omega}{2\pi} \ln \left\{1+ \Gamma_a [(e^{i \lambda} (\cos \phi + \sin \phi) -1) \right. \nonumber\\
&& \left. + (e^{-i \lambda} (\cos \phi - \sin \phi) -1)]\right\} \nonumber\\
&& \times \theta \left(\frac{\Delta_L - |\omega_L|}{\Delta_L}\right) \theta \left(\frac{\Delta_R - |\omega_L|}{\Delta_R}\right) \nonumber\\
&& \times \theta\left(\frac{\Delta_R - |\omega_R|}{\Delta_R}\right) \theta\left(\frac{\Delta_L - |\omega_R|}{\Delta_L}\right), \label{cgfa1}
\end{eqnarray}
where
\begin{eqnarray*}
\Gamma_a = \frac{\Gamma \Delta_R \Delta_L}{(\Delta_L^2 - \omega_L^2)^{1/2}(\Delta_R^2 - \omega_R)^{1/2}},
\end{eqnarray*}
using $\omega_{L/R} = \omega - \mu_{L/R}$.\\
Of course, this result is in perfect accordance with the form in [\onlinecite{PhysRevLett.87.197006}], where, however, the transmission coefficient is slightly different as there the case of a multichannel diffusive contact is treated, whereas here we discuss the case of a single channel ballistic contact.\\
Compared to Refs. [\onlinecite{PhysRevB.68.035105},\onlinecite{PhysRevB.54.7366}] we only observe the first harmonic of the current in the Josephson frequency since we only treat the case of small interface transparency. Indeed, we immediately find for $V< 2 \Delta_{L/R}$
\begin{eqnarray*}
\langle I_{a,1}\rangle (\tau) \propto \sin[\phi(\tau)],
\end{eqnarray*}
which corresponds to the dc- and ac-Josephson current depending on whether a voltage is applied or not.\\
Calculating the noise from Eq. (\ref{cgfa1}) we find that it may become negative even when no voltage is applied. This peculiarity is well known\cite{PhysRevLett.87.197006} and can be traced back to the fact that probabilities $P_q(\phi)$ in Eq.´ (\ref{genfou}) can become negative for SCs due to the additional dependence on the phase difference $\phi$. In this case the interpretation of $P_q(\phi)$ from the Wigner representation as a reasonable probability is rendered impossible. This fact is discussed in more detail in Refs. [\onlinecite{PhysRevLett.87.197006}, \onlinecite{kindermann}].

\section{Multiple Andreev reflections}\label{sectionmar}

We go over to the evaluation of the first part of Eq. (\ref{derivfull})
\begin{eqnarray}
\left \langle \frac{\partial T_{a,2}^{\lambda(t)}}{\partial \lambda}\right\rangle_\lambda &=& - \frac{\gamma^2}{2} \sum_\sigma \int \frac{d\omega}{2\pi} \left[e^{i \lambda} g_{L\sigma}^{-+}(\omega) {\cal G}_{R\sigma}^{\lambda +-}(\omega) \right. \nonumber\\
&& \left. - e^{-i \lambda} g_{L\sigma}^{+-}(\omega) {\cal G}_{R\sigma}^{\lambda-+}(\omega)\right]. \label{deriva2}
\end{eqnarray}
From the second part of Eq. (\ref{derivfull}) we obtained an ac-current contribution for $V\neq 0$. Eq. (\ref{deriva2}) will now produce further dc-current contributions known as MARs and discussed in [\onlinecite{PhysRevLett.74.2110},\onlinecite{PhysRevLett.75.1831}]. Our calculation is closely related to Ref. [\onlinecite{PhysRevB.54.7366}]. We first discuss $V/2 < \Delta_{L/R}$ and will complete our calculation later.\\
The time-dependent coupling in Eq. (\ref{htunnel}) allows for a finite result for Eq. (\ref{deriva2}) even for voltages below the gap. For the following calculation it is easier to introduce combined GFs of the normal and anomalous part
\begin{eqnarray*}
\hat{g}_{\alpha}(\omega) = \left(\begin{array}{cc} g_\alpha(\omega) & f_\alpha(\omega) \\ f_\alpha^+(\omega) & g_\alpha(\omega)\end{array}\right), \; \alpha = L/R.
\end{eqnarray*}
Due to the special time-dependence of the coupling elements every GF admits a Fourier expansion of the form 
\begin{eqnarray*}
\hat{g}_{L/R}(t,t') = \sum_n e^{i n \phi(t')/2} \int \frac{d\omega}{2\pi} e^{-i \omega(t-t')} \hat{g}_{L/R}(\epsilon, \epsilon + neV),
\end{eqnarray*}
which means that $\hat{g}_{L/R}(\epsilon, \epsilon') = \sum_n \hat{g}(\epsilon, \epsilon + neV) \delta(\epsilon - \epsilon' + neV)$. Therefore, in order to calculate the different transport properties we have to find the Fourier components $\hat{g}_{nm,L/R} = \hat{g}_{nm,L/R} (\omega +neV, \omega + meV)$. The Dyson equation for the Fourier components is slightly more complicated as e.g. in the normal-superconductor case\cite{Soller2011425} due to the coupling of different Fourier components. We define $g_{n,L/R} = g_{L,R}(\omega +neV), \; f_{n,L/R} = f_{L,R}(\omega +neV)$ and the Dyson equation can be expressed as\cite{PhysRevB.54.7366} 
\begin{eqnarray*}
\hat{G}_{nm,L/R} &=& \hat{g}_{nm,L/R} \delta_{n,m} + \hat{\epsilon}_{nn} \hat{G}_{nm,L/R} \nonumber\\
&& + \hat{V}_{n,n-2} \hat{G}_{n-2\, m,L/R} + \hat{V}_{n,n+2} \hat{G}_{n+2\,m, L/R},
\end{eqnarray*}
where
\begin{eqnarray*}
\hat{\epsilon}_{nn} &=& \left(\begin{array}{cc} \Sigma_{S,R/L}^n g_{n-1,L/R} & \Sigma_{S,R/L}^{a,n} g_{n+1,L/R} \\ \Sigma_{S,R/L}^{a,n} g_{n-1,L/R} & \Sigma_{S,R/L}^{n} g_{n+1,L/R} \end{array}\right),\\
\hat{V}_{n,n+2} &=& - \left(\begin{array}{cc} \Sigma_{S,R/L}^{a,n} f_{n+1,L/R} & 0 \\ \Sigma_{S,R/L}^{n} f_{n+1,R/L} & 0 \end{array}\right),\\
\hat{V}_{n,n-2} &=& - \left(\begin{array}{cc} 0 & \Sigma_{S,R/L}^n f_{n-1,L/R} \\ 0 & \Sigma_{S,R/L}^{a,n} f_{n-1,L/R} \end{array}\right).
\end{eqnarray*}
We have used the abbreviations
\begin{eqnarray*}
\Sigma_{S,L/R}^n &=& \gamma^2 \left(\begin{array}{cc} g_{n,L/R}^{--} & - e^{-i \lambda} g_{n,L/R}^{-+} \\ -e^{i \lambda} g_{n,L/R}^{+-} & g_{n,L/R}^{++} \end{array}\right),\\
\Sigma_{S,L/R}^{a,n} &=& \gamma^2 \left(\begin{array}{cc} f_{n,L/R}^{--} & - e^{-i \lambda} f_{n,L/R}^{-+} \\ - e^{i \lambda} f_{n,L/R}^{+-} & f_{n,L/R}^{++}\end{array}\right).
\end{eqnarray*}
In the case of low transmission the electronic transport can be described by a sequential tunneling picture\cite{PhysRevB.55.R6137,PhysRevB.54.7366} with transmission coefficients given by a product of tunneling rates for each MAR. The corresponding transmission coefficients are given by
\begin{eqnarray*}
&& \Gamma_{a,n,L/R} = 4 \Gamma^n \prod_{k=1}^{\left\lceil \frac{n-1}{2} \right \rceil} \frac{\Delta_{L/R}^2}{\Delta_{L/R}^2 - (\omega \mp keV/2)^2} \nonumber\\
&& \times \theta \left(\frac{\Delta_{L/R} - |\omega \mp keV/2|}{\Delta_{L/R}}\right) \prod_{k=1}^{\left \lfloor \frac{n-1}{2} \right \rfloor} \frac{\Delta_{R/L}^2}{\Delta_{R/L}^2 - (\omega \pm keV/2)^2} \\
&& \times \theta \left(\frac{\Delta_{R/L} - |\omega \pm keV/2|}{\Delta_{R/L}}\right) \nonumber\\
&& \times \frac{|\omega - neV/2| |\omega + neV/2|}{\sqrt{(\omega + neV/2)^2 - \Delta_R^2} \sqrt{(\omega - neV/2)^2 - \Delta_L^2}} \nonumber\\
&& \times \theta \left(\frac{|\omega + neV/2| - \Delta_R}{\Delta_R}\right) \theta \left(\frac{|\omega - neV/2| - \Delta_L}{\Delta_L}\right).
\end{eqnarray*}
The result for the CGF for $V < 2\Delta_{L/R}$ is then
\begin{eqnarray}
&&\ln \chi_{a,2}(\lambda, \tau) = \tau \int \frac{d\omega}{2\pi} \ln \left\{1+ \sum_{n=2, \; \alpha = L/R}^\infty \Gamma_{a,n,\alpha} \right. \nonumber\\
&& \times [(e^{i \lambda n} -1) n_F(\omega - nV/2) (1- n_F(\omega + nV/2)) \nonumber\\
&& \left. + e^{-i \lambda n} n_F(\omega + nV/2) (1- n_F(\omega - nV/2))]\right\}, \label{cgfmar}
\end{eqnarray}
where $n_F(\omega)$ is the Fermi function.\\
Of course, the results for the current and noise that can be obtained from Eq. (\ref{cgfmar}) are identical to the ones obtained in [\onlinecite{PhysRevB.54.7366}, \onlinecite{PhysRevLett.82.4086}] for tunnel junctions.

\section{Andreev reflections for intermediate voltages} \label{sectionintermediate}

So far we have considered $V< 2 \Delta_{L,R}$ in which case both SCs contribute to the MARs. However, for $\Delta_L \neq \Delta_R$ there is an intermediate regime in which only one of the SCs contributes to the AR processes. The description of this intermediate regime is rather involved, however, it does not lead to surprising dependencies on the counting fields. We consider, without loss of generality, $\Delta_L \geq \Delta_R$ and the following regimes for the energies
\begin{itemize}
\item b: $\Delta_R < |\omega + V/2 | < \Delta_L, \; |\omega - V/2|< \Delta_{L/R}$,
\item c: $\Delta_R < |\omega + V/2 | < \Delta_L, \; \Delta_R < |\omega - V/2| < \Delta_L$,
\item d: $\Delta_R < |\omega + V/2 | < \Delta_L, \; |\omega - V/2| > \Delta_{L/R}$,
\item e: $|\omega + V/2 < \Delta_{L/R}, \; \Delta_R < |\omega - V/2| < \Delta_L$,
\item f: $|\omega + V/2| > \Delta_{L/R}, \; \Delta_R < |\omega - V/2| < \Delta_L$,
\item g: $|\omega - V/2| < \Delta_{L/R}, \; |\omega + V/2| > \Delta_{L/R}$,
\item h: $|\omega + V/2| < \Delta_{L/R}, \; |\omega - V/2| > \Delta_{L/R}$,
\item i: $|\omega - V/2| > \Delta_{L/R}, \; |\omega + V/2| > \Delta_{L/R}$.
\end{itemize}
We considered each configuration and solved the Dyson series. Using Eq. (\ref{derivfull}) we obtained the following results
\begin{widetext}
\begin{eqnarray*}
\ln \chi_c(\lambda, \tau) &=& \tau \int \frac{d\omega}{2\pi} \ln \left\{1+ 4 \Gamma_c [n_F(\omega - V) (1-n_F(\omega + V)) (e^{2i \lambda} -1) + n_F(\omega + V) (1- n_F(\omega - V))(e^{- 2i \lambda} -1)]\right\} \nonumber\\
&& \theta \left(\frac{|\omega_L| - \Delta_R}{\Delta_R}\right) \theta \left(\frac{\Delta_L - |\omega_L|}{\Delta_L}\right) \theta \left(\frac{|\omega_R| - \Delta_R}{\Delta_R}\right) \theta \left(\frac{\Delta_L - |\omega_R|}{\Delta_L}\right),\\
\ln \chi_d(\lambda, \tau) &=& 2 \tau \int \frac{d\omega}{2\pi} \ln \left\{ 1+ \Gamma_d[n_L(1- n_R) (e^{i \lambda} -1) + n_R(1- n_L) (e^{-i \lambda} -1)]\right\} \nonumber\\
&& \times \theta \left(\frac{|\omega_L| - \Delta_L}{\Delta_L}\right) \theta \left(\frac{\Delta_L - |\omega_R|}{\Delta_L}\right) \theta \left(\frac{|\omega_R| - \Delta_R}{\Delta_R}\right),\\
\ln \chi_e(\lambda, \tau) &=& 2 \tau \int \frac{d\omega}{2\pi} \ln \left\{1+ \Gamma_a [(e^{i \lambda} -1)(\cos \phi + \sin \phi) + (e^{-i \lambda} -1) (\cos \phi - \sin \phi)]\right\} \nonumber\\
&& \times \theta \left(\frac{\Delta_L - |\omega_L|}{\Delta_L}\right) \theta \left(\frac{|\omega_L| - \Delta_R}{\Delta_R}\right) \theta \left(\frac{\Delta_R - |\omega_R|}{\Delta_R}\right),\\
\ln \chi_f(\lambda, \tau) &=& 2 \tau \int \frac{d\omega}{2\pi} \ln \left\{1 + 4 \Gamma_{f} [n_F(\omega - V) (1- n_F(\omega + V))(e^{2i \lambda} -1) + n_F(\omega + V) (1- n_F(\omega - V)) (e^{-2i \lambda} -1)]\right\}\nonumber\\
&& \times \theta \left(\frac{|\omega_R| - \Delta_L}{\Delta_L}\right) \theta \left(\frac{|\omega_L| - \Delta_R}{\Delta_R}\right) 
\theta \left(\frac{\Delta_L - |\omega_L|}{\Delta_L}\right),\\
\ln \chi_i(\lambda, \tau) &=& 2 \tau \int \frac{d\omega}{2\pi} \ln \left\{1+ 4 \Gamma_i [n_L(1-n_R) (e^{i \lambda} -1) + n_R(1-n_L) (e^{-i \lambda} -1)]\right\}\nonumber\\
&& \times \theta \left(\frac{|\omega_L| - \Delta_L}{\Delta_L}\right) \theta \left(\frac{|\omega_R| - \Delta_L}{\Delta_L}\right),
\end{eqnarray*}
\end{widetext}
where we have used $n_{L/R} = n_F(\omega - \mu_{L/R})$ and
\begin{eqnarray*}
\Gamma_c &=& \frac{\Gamma^2 \Delta_L^2 |\omega_R|^2}{(\Delta_L^2 - \omega_L^2) (\omega_R^2 - \Delta_R)^{1/2} (\omega_L^2 - \Delta_R^2)^{1/2}} = \Gamma_{f},\\
\Gamma_d &=& \frac{\Gamma |\omega_L| |\omega_R|}{2 (\omega_L^2 - \Delta_L^2)^{1/2} (\omega_R^2 - \Delta_R^2)^{1/2}} = \Gamma_i.
\end{eqnarray*}
The other energy configurations do not contribute to charge transfer. We observe an interplay of AR and single-electron transmission with the typical dependencies on the counting fields. In this respect no new noise features apart from those already discussed in the literature\cite{PhysRevLett.82.4086} appear.\\
The full result for the CGF for a superconducting tunnel junction have the form
\begin{eqnarray}
\ln \chi(\lambda, \phi, \tau) &=& \ln \chi_{a,1}(\lambda, \phi, \tau) + \ln \chi_{a,2}(\lambda, \tau) + \ln \chi_c(\lambda, \tau) \nonumber\\
&& + \ln \chi_d(\lambda, \tau) + \ln \chi_e(\lambda, \phi, \tau) + \ln \chi_f(\lambda, \tau) \nonumber\\
&& + \ln \chi_i (\lambda, \tau). \label{fullcgf}
\end{eqnarray}
Of course, choosing $\Delta_R = 0$ we reproduce the result in Ref. [\onlinecite{Soller2011425}] in the low transparency limit.\\
Nonetheless, we want to show that the above derived expressions indeed allow for a correct description of a superconducting tunnel junction at finite but possibly different $\Delta_{L/R}$.

\section{Zero temperature SIS junction} \label{sectionexp}

\begin{figure}[ht]
\includegraphics[width=0.7\columnwidth]{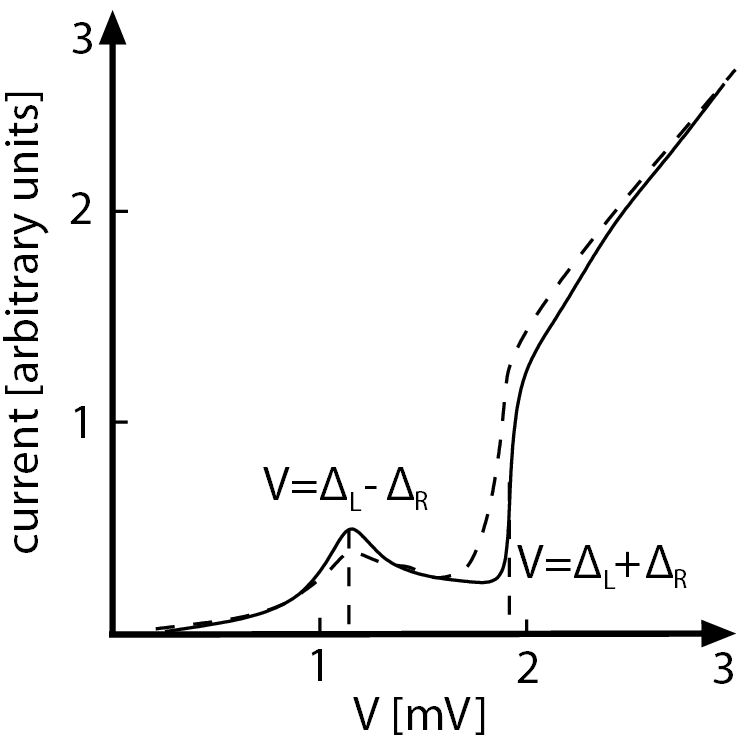}
\caption{Current in a Nb-Sn junction at $T=3.38$K with $\Delta_{\mbox{Sn}} =0.19 \cdot 10^{-3}$ eV, $\Delta_{\mbox{Nb}} = 1.49 \cdot 10^{-3}$ eV. The experimental measurement from [\onlinecite{PhysRev.128.591}] is shown as the dashed curve and compared to the theoretical prediction using Eq. (\ref{siscurrent}), solid curve.}
\label{fig2}
\end{figure}

We want to discuss how to reproduce the results in [\onlinecite{tinkham}] from the semiconductor model. We limit ourselves to $T=0$ and in the first part only consider $\Delta_L = \Delta_R = \Delta$. Furthermore we limit ourselves to the lowest order in $\Gamma$ regardless of the energy dependence so that the only contributing parts are $\ln \chi_{a,1}$ and $\ln \chi_i$. This approximation is typically justified for the experimental situation of two superconductors coupled via an insulating barrier (SIS junction). We disregard the ac-component and are consequently left with the calculation of the current from $\ln \chi_i$. We arrive at the result also presented in [\onlinecite{tinkham}]
\begin{widetext}
\begin{eqnarray*}
\langle I\rangle_1 = \left\{ \begin{array}{c} \theta\left(\frac{V- 2\Delta}{\Delta}\right) 8 \Gamma \int_{\Delta}^{V- \Delta} \frac{d\omega}{2\pi} \frac{\omega (V- \omega)}{(\omega^2 - \Delta^2) [(V- \omega)^2 - \Delta^2]^{1/2}}, \; V> 0 \\ - \theta\left(\frac{|V| - 2\Delta}{\Delta}\right) 8 \Gamma \int_{\Delta}^{|V|- \Delta} \frac{d\omega}{2\pi} \frac{\omega (|V| - \omega)}{(\omega^2 - \Delta^2) [(|V|- \omega)^2 - \Delta^2]^{1/2}}, \; V< 0 \end{array}\right..
\end{eqnarray*}
Since the two cases only differ in sign we choose $V>0$ from now on. Using elliptic integrals and reintroducing SI-units we arrive at
\begin{eqnarray*}
\langle I\rangle_1 = 2 \frac{G_0}{e} \theta\left(\frac{eV - 2\Delta}{\Delta}\right) \left\{\frac{(eV)^2}{eV + 2 \Delta} K(\alpha) - (eV + 2 \Delta) [K(\alpha) - E(\alpha)]\right\},
\end{eqnarray*}
\end{widetext}
where $\alpha = \frac{eV - 2 \Delta}{eV + 2 \Delta}$ and $\tilde{T} = 4 \gamma^2 \pi^2 \rho_{0L} \rho_{0R}$. Using $K(0) = E(0) = \pi /2$ one obtains the well-known discontinuity in the tunneling current at $eV = 2 \Delta$ to $\pi/4$ times the tunneling current in the case of normal contacts.\\
The case of $\Delta_L \neq \Delta_R$ cannot be treated integrated analytically anymore. We obtain for the current to lowest order in $\Gamma$
\begin{eqnarray}
\langle I \rangle_2 &=& 16 \int \frac{d\omega}{2\pi} \left\{\frac{\Gamma |\omega_R| |\omega_L|}{2(\omega_R^2 - \Delta_R^2)^{1/2}(\omega_L^2 - \Delta_L^2)^{1/2}} \right. \nonumber\\
&& \left. \times [n_L(1-n_R) + n_R(1-n_L)]\right\} \nonumber\\
&& \times \theta\left(\frac{|\omega_L| - \Delta_L}{\Delta_L}\right) \left(\frac{|\omega_R| - \Delta_R}{\Delta_R}\right). \label{siscurrent}
\end{eqnarray}
We compare this result to the experimental data from [\onlinecite{PhysRev.128.591}] for a Nb-Sn junction in Fig. \ref{fig2}.
\\
A cusp is observed for the current at $V=|\Delta_L - \Delta_R|$ and a steep rise is again observed at $|V| = \Delta_L + \Delta_R$. For $|V| = |\Delta_L - \Delta_R|$ the DOS of electrons participating in the charge transfer in both SCs is maximal and allows for a calculation of the order parameters.\\
We do not observe perfect agreement as the surfaces of the SCs in the SIS junction can be nonhomogeneous leading to different characteristics of the SC at different positions. Furthermore Nb is already a strongly coupled SC making BCS theory not perfectly applicable. Nonetheless, the agreement is accetable and shows that the CGF in Eq. (\ref{fullcgf}) allows for a description of the case $\Delta_L \neq \Delta_R$.

\section{Conclusions} \label{sectionconclude}

To conclude, we have derived the CGF for a superconducting tunnel junction taking multiple Andreev reflections, Josephson tunneling and possibly different gaps of the two superconductors into account. In this way our result generalizes previous results for the CGF taking only certain aspects into account. The understanding of the CGF in simple tunnel junctions paves the way to understanding more involved geometries.\cite{PhysRevB.55.R6137}\\
The author would like to thank A. Levy Yeyati, A. Komnik, S. Maier, J.C. Cuevas and D. Breyel for many interesting discussions.

\end{document}